\documentclass{article}
\usepackage[T1]{fontenc}
\usepackage[utf8]{inputenc}
\usepackage{ismir} 
\usepackage{amsmath,amssymb,cite,url}
\usepackage{graphicx}
\usepackage{color}
\usepackage{multirow}
\usepackage{tikz}

\newcommand{\rhobar}[1]{%
  \begin{tikzpicture}[baseline=-2pt, x=8mm, y=2.6pt]
    \path[use as bounding box] (-0.76, -1.6) rectangle (0.76, 1.6);
    \pgfmathsetmacro{\rhoval}{#1}%
    \pgfmathtruncatemacro{\rhoshade}{18 + 50*abs(\rhoval)}%
    \ifdim \rhoval pt > 0pt
      \fill[red!\rhoshade]  (0,-1) rectangle (\rhoval,1);
      \node[anchor=west, font=\tiny, inner sep=0pt]
            at (0.045,0) {$#1$};
    \else
      \fill[blue!\rhoshade] (\rhoval,-1) rectangle (0,1);
      \node[anchor=east, font=\tiny, inner sep=0pt]
            at (-0.045,0) {$#1$};
    \fi
    \draw[black!55, line width=0.18pt] (0,-1.6) -- (0,1.6);
  \end{tikzpicture}%
}

\title{\scalebox{0.974}[1.0]{\mbox{Do Music Generative Models Understand Musical Qualities?}}\\ Automatic Music Evaluation with Model-Intrinsic Signals}

\threeauthors
  {Xiaosha Li} {Georgia Institute of Technology \\ \texttt{xiaosha@gatech.edu}}
  {Chun Liu} {ByteDance Inc. \\ \texttt{chun.liu@bytedance.com}}
  {Ziyu Wang} {New York University $\cdot$ MBZUAI \\ \texttt{ziyu.wang@nyu.edu}}

\def\authorname{X. Li, C. Liu, and Z. Wang}

\usepackage[bookmarks=false,pdfauthor={\authorname},pdfsubject={\pdfsubject},hidelinks]{hyperref}

\begin{document}

\maketitle

\begin{abstract}
Current music generative models can produce high-quality music, but does this ability imply that they ``understand'' the musical qualities of their outputs, and is that understanding aligned with human evaluation? Previous attempts to use the likelihood of a generative model to evaluate music, an approach commonly used in text, have proven unsuccessful, leading researchers to rely on standalone supervised music evaluation models. In this paper, we answer this question affirmatively: we show that a model's intrinsic signals---derived from its hidden representations and predictions---are strongly correlated with human ratings. In particular, we study MusicGen and consider three types of features: (1) prediction loss, (2) prediction entropy, and (3) concepts extracted from the model using a sparse autoencoder (SAE). Using these features, we train a lightweight prediction model to estimate subjective ratings. We evaluate these features both individually and in combination. We hypothesize that these signals parallel the listening process: the temporal and frequency-domain structure of loss and entropy reflects listeners' expectation and surprise, while gradient directions in SAE latent space predict perceived quality. Experiments on five human-evaluation benchmarks spanning continuous ratings and pairwise preferences confirm this hypothesis, with SAE latents carrying most of the predictive signal.\footnote{We release the code (\url{https://github.com/YoEv/MEva}), an audio demo page (\url{https://yoev.github.io/MEva}), and the evaluator checkpoints (\url{https://huggingface.co/Dev4IC/MEva-checkpoints}).}
\end{abstract}

\section{Introduction}\label{sec:introduction}

Generative music models have advanced rapidly, yet judging the musical qualities~\cite{justus2002music,koelsch2011toward} of their outputs without human listeners remains an open problem.  Recent years have seen a growing body of human-evaluation datasets for music generation---MusicEval~\cite{liu2025musiceval}, SongEval~\cite{yao2025songeval}, MusicPref~\cite{huang2025musicpref}, AIME~\cite{grotschla2025aime}, and Music Arena~\cite{kim2025musicarena}---and of audio-based verifiers that attempt to approximate these ratings, most notably Meta Audiobox Aesthetics~\cite{tjandra2025audiobox} and the music-instruction benchmarks CMI-Bench~\cite{ma2025cmibench} and its reward-model variant~\cite{ma2026cmireward}.  These approaches share a common assumption: musical qualities are inferred solely from the generated audio, treating the music model as a black box. Such systems automate the rating step but leave the deeper question untouched: \emph{why} does a human listener rate one clip highly and another poorly? Aesthetics research suggests no general answer exists; what can be measured is rating behavior in controlled settings.

Music cognition offers a more principled vantage point. Human appreciation has long been modeled as two exposure-driven processes: repeated exposure shapes preference toward the familiar (the Mere Exposure Effect~\cite{zajonc1968mereexposure,schellenberg2008exposure}), and real-time listening continually forms and re-evaluates predictions, with surprises such as the deceptive cadence carrying affective charge~\cite{meyer1956emotion,huron2006sweet}. Under this model of cognition---one among several---judging music reads as an alignment between the listener's exposure-shaped representations and expectations and the audio they encounter.

An autoregressive generative music model parallels parts of this process. It is trained on large amounts of music data, and its next-token logits encode an expectation over what the music should do next. This leads us to the central question of this paper: \emph{can a generative music model judge how highly a human would rate its own output, using only its own intrinsic signals---without ever leaving its internal state to consult an external audio verifier?}  The broader version is just as natural: a generation model that can \emph{generate} highly rated music---can it also \emph{evaluate} it in its own traces?

We answer this question empirically. We extract three intrinsic signals from a single pretrained autoregressive music generator---per-token loss, next-token predictive entropy, and sparse-autoencoder (SAE) latents of the hidden states---and align them to human ratings from five diverse benchmarks (MusicEval, SongEval, MusicPref, AIME, Music Arena) under one unified protocol.  Empirically, the more often the model is confident yet wrong (low entropy, high loss), the lower the human rating; SAE latents independently organize clips along a quality axis without human labels. Gemini-2.5-Flash audio captioning corroborates both findings, with a professional music producer agreeing on the latent concepts.  Concretely, our contributions are:
\pagebreak
\begin{itemize}
  \item \textbf{Intrinsic-signal alignment framework.}
    Model-internal signals from a single pretrained music generator
    align with human ratings across five benchmarks under one
    protocol.
  \item \textbf{Loss--entropy mismatch as a low-rating signature.}
    The mismatch isolates an \emph{overconfident-wrong} configuration
    that is the dominant local correlate of low ratings, corroborated
    by audio captioning.
  \item \textbf{Representational evidence from SAE.}  SAE latents,
    trained without ratings, separate highly rated clips from poorly rated clips, corroborated by audio captioning and a professional
    producer.
\end{itemize}

\section{Related Work}\label{sec:background}

We connect three strands of music evaluation---audio-based verifiers, human-rated benchmarks, and the cognitive science of preference---to a fourth largely missing for music: model-internal signals (loss, entropy, sparse representations) standard in language modeling and interpretability.

Several recent benchmarks score generated music directly from the output waveform. Audiobox Aesthetics~\cite{tjandra2025audiobox} evaluates generated music along two complementary perspectives: production-side quality (Production Complexity, Production Quality) and content-side perception (Content Enjoyment, Content Usefulness). CMI-Bench~\cite{ma2025cmibench} reframes traditional MIR annotations as music instruction-following tasks for audio-text LLMs, while CMI-RewardBench~\cite{ma2026cmireward} introduces a reward-modeling benchmark for generated music across musicality, text--music alignment, and compositional multimodal instruction alignment.

Recent benchmarks span expert continuous ratings on text-to-music clips (MusicEval~\cite{liu2025musiceval}) and full songs (SongEval~\cite{yao2025songeval}); pairwise head-to-head battles (MusicPref~\cite{huang2025musicpref}, Music Arena~\cite{kim2025musicarena}, the latter analogous to Chatbot Arena~\cite{chiang2024chatbotarena}); and the survey-style AIME~\cite{grotschla2025aime}, in which each clip appears in twelve pairwise comparisons per rating axis. Their varied supervision formats (MOS, multi-axis
ratings, pairwise outcomes) are unified in Section~\ref{sec:experiments}. Both families judge from the waveform alone, leaving the \emph{why} of preference unaddressed. In the survey of Lerch et al.~\cite{lerch2025survey}, our method falls under objective, reference-free quality prediction trained on human ratings---differing from existing predictors in taking as input the generator's internal signals rather than audio embeddings.

Of the many mechanisms music-cognition research proposes, two map naturally onto an autoregressive model: repeated exposure builds preference-shaping representations of timbre, tonality, and structure (the Mere Exposure Effect~\cite{zajonc1968mereexposure,schellenberg2008exposure}), and real-time listening generates expectations whose violations carry affective weight~\cite{meyer1956emotion,huron2006sweet}. A pretrained autoregressive music model instantiates both: its hidden states encode representations shaped by training exposure~\cite{singh2025steering,muellereberstein2023subspace}, and its
next-token logits are a learned expectation~\cite{copet2023musicgen}.

If judgments of musical qualities rest on prediction and familiarity, the natural place to read them off is the generator's own internal signals. Li et al.~\cite{li2026noiseloss} show that, in music LLMs, mean loss does not align with human-rated musicality and local minima do not align with preferred regions; what does carry signal is the shape of the loss curve over time. Parallel work on language models shows that perplexity (and therefore negative log-likelihood) cannot by itself distinguish confidently correct from confidently incorrect predictions~\cite{velickovic2026perplexity}. This aligns with a long-standing observation in the calibration literature: modern neural networks tend to be
overconfident~\cite{guo2017calibration}, and penalizing overconfident output distributions acts as an implicit regularizer~\cite{pereyra2017confidence}.

Entropy thus enters naturally as a complementary signal. It has been used as a reliability indicator at decoding time~\cite{qiu2024entropy}. At the token level, per-token loss outperforms entropy as an error indicator~\cite{malinin2021uncertainty}; at the sequence level, the average predictive entropy of the softmax outperforms mean loss as a quality indicator~\cite{fomicheva2020unsupervised}. The two are therefore complementary at different temporal scales, motivating an encoder that consumes both curves jointly.

Sparse autoencoders (SAEs)~\cite{cunningham2023sparse, bricken2023monosemanticity, templeton2024scaling} map dense hidden activations to a sparser latent space, decomposing entangled representations into more localized, interpretable units. Singh et al.~\cite{singh2025steering} apply this to a generative music model and recover sparse latents whose activations correspond to specific, musically meaningful concepts; using them as inference-time steering vectors yields controllable changes in generated audio. We reuse the same latent readout but repurpose it for evaluation, asking whether the latents that Singh et al.\ interpret concept-by-concept also organize clips along an axis aligned with human ratings. Unlike loss and entropy---summaries of the output distribution---%
SAE features probe the model's internal representation, explained in
Section~\ref{subsec:signals}.

\begin{figure}[t]
  \centering
  \includegraphics[width=\linewidth]{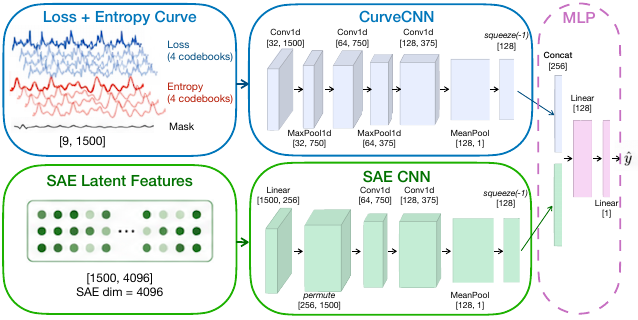}
  \caption{Hybrid prediction model: one 1D-CNN encoder per intrinsic
    signal, late concatenation (Eq.~\ref{eq:hybrid_concat}), and a
    shared MLP head producing $\hat y\in[1,5]$.}
  \label{fig:architectures}
\end{figure}

\begin{figure*}[t]
  \centering
  \includegraphics[width=0.95\linewidth]{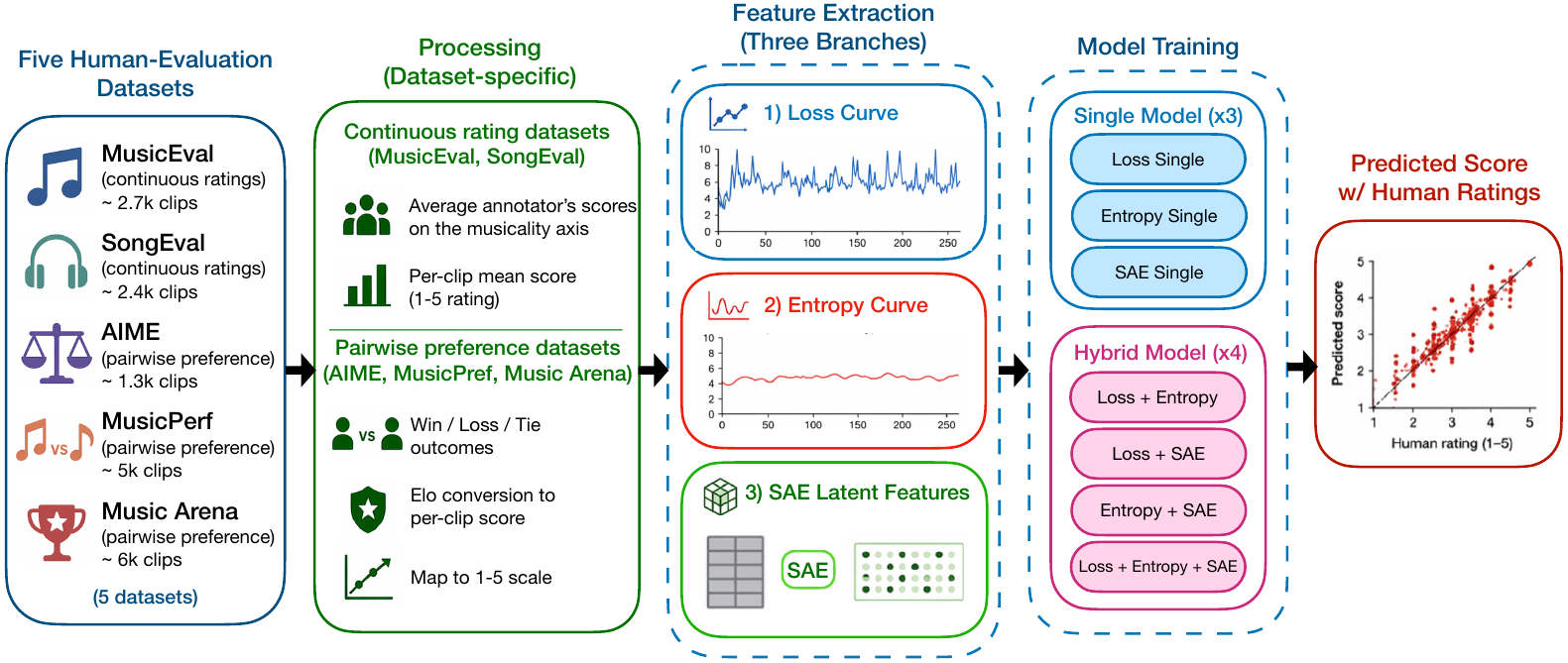}
  \caption{Experimental flowchart: human-evaluation data, per-clip
    loss / entropy / SAE features, and the hybrid network with its
    ablations, evaluated against human ratings.}
  \label{fig:flowchart_models}
\end{figure*}

\section{Methods}\label{sec:methods}

From a pretrained autoregressive music model we extract three intrinsic signals---token-level loss $\ell_t$, predictive entropy $H_t$, and sparse-autoencoder (SAE) latents $z_t$---each chosen to mirror, at the level of the model, one of the listener-side processes invoked in Section~\ref{sec:introduction}, and train a single neural network that maps them to a human rating.

\subsection{Model-Intrinsic Signals}
\label{subsec:signals}

Given a generator $p_\theta$ and a token sequence $\{x_t\}_{t=1}^{T}$, we extract three complementary per-step signals, all computed under teacher-forcing on held-out audio.

\textbf{Loss curve.}
Similar to a listener's \emph{surprise}, the loss $\ell_t$ in the curve $\mathbf{l}$ is the negative log-likelihood of the ground-truth token under the model's next-token distribution:
\begin{equation}
\begin{aligned}
\ell_t &= -\log p_\theta(x_t \mid x_{<t}), \\
\mathbf{l} &= [\ell_1,\ldots,\ell_T].
\end{aligned}
\end{equation}

\textbf{Predictive entropy curve.}
Like how uncertain a listener is about what comes next, $H_t$ in the curve $\mathbf{h}$ is the Shannon entropy of $p_\theta(\cdot\mid x_{<t})$ over a vocabulary of size $K$---low when the model concentrates on a few continuations, high when it spreads across many:
\begin{equation}
\begin{gathered}
H_t = -\sum_{k=1}^{K} p_\theta(k\mid x_{<t}) \log p_\theta(k\mid x_{<t}), \\
\mathbf{h} = [H_1,\ldots,H_T].
\end{gathered}
\end{equation}

\textbf{SAE latents.}
Where $\ell_t$ and $H_t$ summarize only the output distribution, $z_t$ probes the model's internal state and---mirroring the disentangled concepts (timbre, register, harmony) a listener acquires from exposure---re-expresses each hidden activation as a sparse combination of latent features that tend to align with a single interpretable musical concept. For a hidden activation $u_t\in\mathbb{R}^{d_{\text{in}}}$, the SAE produces sparse codes $z_t=f_{\text{enc}}(u_t)\in\mathbb{R}^{d_{\text{sae}}}$ ($d_{\text{sae}}>d_{\text{in}}$) and reconstructions $\hat{u}_t = f_{\text{dec}}(z_t)$, trained by Singh et al.\ with the reconstruction-plus-$\ell_1$ objective. Because only a few entries of $z_t$ are active per step, the SAE acts as an overcomplete sparse dictionary that turns dense polysemantic activations into a sparse combination of more disentangled, concept-aligned features~\cite{singh2025steering}. We reuse the music-SAE of Singh et al.~\cite{singh2025steering} without retraining.

\subsection{Prediction Model}
\label{subsec:architectures}

Our main neural network is a three-branch \emph{hybrid} architecture that consumes all three intrinsic signals together (Figure~\ref{fig:architectures}). Each signal $s\in\{\ell, H, z\}$ is fed to a dedicated 1D convolutional encoder $f_s$ that produces a per-branch representation $r_s\in\mathbb{R}^{d_r}$. The three branch representations are combined by late concatenation,
\begin{equation}
r = [r_{\ell};\, r_H;\, r_z],
\label{eq:hybrid_concat}
\end{equation}
and a shared Multi-Layer Perceptron (MLP) head maps $r$ to the predicted clip-level rating
$\hat y\in[1,5]$. Single- and two-signal ablations drop branches accordingly; encoder sizes and training are in Section~\ref{subsec:exp_training}.

\begin{table*}[t]
  \centering
  \scriptsize
  \setlength{\tabcolsep}{2.5pt}
  \begin{tabular}{ll cc cc cc cc cc cc}
    \hline
    \multirow{2}{*}{Group} & \multirow{2}{*}{Model}
      & \multicolumn{2}{c}{MusicEval}
      & \multicolumn{2}{c}{SongEval}
      & \multicolumn{2}{c}{AIME}
      & \multicolumn{2}{c}{MusicPref}
      & \multicolumn{2}{c}{Music Arena}
      & \multicolumn{2}{c}{All benchmarks} \\
    \cline{3-4} \cline{5-6} \cline{7-8} \cline{9-10} \cline{11-12} \cline{13-14}
    & & $r$ & $\rho$ & $r$ & $\rho$ & $r$ & $\rho$
        & $r$ & $\rho$ & $r$ & $\rho$ & $r$ & $\rho$ \\
    \hline
    \multirow{5}{*}{Baseline}
      & Aesthetics-CE
        & 0.62 & 0.64 & 0.57 & 0.57 & 0.41 & 0.38 & 0.04 & 0.03 & 0.24 & 0.26 & 0.20 & 0.21 \\
      & Aesthetics-CU
        & 0.62 & 0.64 & 0.63 & 0.70 & 0.48 & 0.45 & 0.04 & 0.04 & 0.22 & 0.24 & 0.18 & 0.19 \\
      & Aesthetics-PC
        & 0.08 & 0.04 & 0.32 & 0.27 & 0.01 & 0.01 & 0.01 & 0.00 & 0.15 & 0.23 & 0.07 & 0.09 \\
      & Aesthetics-PQ
        & 0.56 & 0.59 & 0.61 & 0.65 & 0.42 & 0.40 & 0.03 & 0.03 & 0.33 & 0.40 & 0.20 & 0.23 \\
      & Mean Loss
        & -0.02 & -0.07 & 0.24 & 0.11 & -0.10 & -0.09 & -0.30 & -0.41 & -0.08 & -0.07 & -0.08 & -0.13 \\
    \hline
    \multirow{3}{*}{Single}
      & \textsc{Single-Loss}
        & 0.55/0.55 & 0.56/0.55 & 0.60/0.58 & 0.61/0.61 & 0.58/0.62 & 0.56/0.60 & \textbf{0.95}/0.95 & \textbf{0.90}/0.89 & 0.66/0.63 & 0.71/0.71 & 0.64/0.67 & 0.62/0.65 \\
      & \textsc{Single-Entropy}
        & 0.66/0.63 & 0.68/0.65 & 0.66/0.72 & 0.66/0.73 & 0.69/0.66 & 0.68/0.62 & \underline{0.94}/0.95 & 0.89/0.90 & 0.66/0.63 & 0.74/0.70 & 0.70/0.69 & 0.70/0.67 \\
      & \textsc{Single-SAE}
        & \underline{0.79}/\underline{0.77} & \underline{0.80}/\underline{0.79} & 0.85/\underline{0.85} & 0.85/\underline{0.84} & 0.82/\underline{0.85} & 0.82/\underline{0.85} & 0.78/0.98 & 0.71/0.93 & \underline{0.78}/\underline{0.79} & \underline{0.84}/0.84 & 0.73/0.81 & 0.73/\underline{0.80} \\
    \hline
    \multirow{4}{*}{Hybrid}
      & \textsc{Loss+Entropy}
        & 0.59/0.57 & 0.61/0.59 & 0.69/0.65 & 0.70/0.67 & 0.67/0.67 & 0.63/0.64 & \underline{0.94}/0.96 & \underline{0.90}/0.91 & 0.67/0.66 & 0.74/0.73 & 0.63/0.69 & 0.62/0.67 \\
      & \textsc{Loss+SAE}
        & \underline{0.78}/\textbf{0.79} & \underline{0.80}/\textbf{0.80} & 0.85/\textbf{0.86} & 0.83/\underline{0.85} & \underline{0.84}/\textbf{0.86} & \underline{0.84}/\textbf{0.87} & \underline{0.94}/\underline{0.99} & 0.86/\underline{0.94} & \underline{0.78}/\textbf{0.80} & \underline{0.83}/\textbf{0.86} & 0.79/\underline{0.82} & \underline{0.78}/\underline{0.81} \\
      & \textsc{Entropy+SAE}
        & \underline{0.78}/\underline{0.78} & \underline{0.79}/\underline{0.79} & \textbf{0.87}/\underline{0.86} & \textbf{0.87}/\textbf{0.85} & \textbf{0.85}/\underline{0.85} & \textbf{0.85}/\underline{0.85} & 0.89/0.99 & 0.85/\underline{0.94} & 0.77/\underline{0.79} & \underline{0.84}/0.85 & 0.78/\textbf{0.83} & \underline{0.79}/\textbf{0.81} \\
      & \textsc{Loss+Entropy+SAE}
        & \textbf{0.79}/\underline{0.77} & \textbf{0.80}/\underline{0.79} & 0.84/0.84 & 0.83/0.83 & \underline{0.82}/\underline{0.85} & 0.81/0.84 & \underline{0.94}/\textbf{0.99} & \underline{0.89}/\textbf{0.94} & \textbf{0.79}/\underline{0.79} & \textbf{0.84}/0.85 & \textbf{0.81}/\underline{0.82} & \textbf{0.80}/\underline{0.80} \\
    \hline
  \end{tabular}
  \caption{Pearson $r$ and Spearman $\rho$ vs.\ human evaluation;
    trained cells are MusicGen-small\,/\,large. Trained rows: held-out
    test split per benchmark; Audiobox rows: zero-shot on the full set;
    \emph{All benchmarks}: union of the five held-out splits. Per size, \textbf{bold} denotes the best score in each column, and \underline{underlining} denotes scores tied with it
 (bootstrap, $p{\geq}0.05$~\cite{demsar2006}).}
  \label{tab:main_results}
\end{table*}

\subsection{Training Objective Across Datasets}
\label{subsec:objective}

We train all variants of the neural network against a single per-clip target $y_i\in[1,5]$ with one mean-squared-error objective,
\begin{equation}
\mathcal{L}_{\text{reg}}
 = \frac{1}{N}\sum_{i=1}^{N}(\hat y_i - y_i)^2,
\label{eq:lreg}
\end{equation}
regardless of supervision format. MusicEval and SongEval supply $y_i$ directly as the per-clip average expert rating. MusicPref, AIME and Music Arena release only pairwise win/loss/tie outcomes, which we convert to per-clip continuous ratings with the Elo scheme~\cite{elo1978rating}: each competitor's scalar rating is updated after every match by an amount proportional to the gap between observed and rating-implied outcomes, and iterating across all matches yields a system-level continuous ranking analogous to chess. This is the same kind of system-level ranking AIME~\cite{grotschla2025aime} and the Chatbot-Arena-style Music Arena~\cite{kim2025musicarena} also report across music generation systems---a property the standard alternatives do not share: pairwise margin (hinge/ReLU)~\cite{herbrich1999support} and RankNet~\cite{burges2005learning} losses are training objectives rather than data conversions, and Bradley--Terry~\cite{bradley1952rank} fits the same competitor-strength likelihood as Elo but only as a batch MLE, with no per-match increment of the kind we use to assign
each clip its own rating from a single battle. Per-dataset details are in Section~\ref{subsec:exp_data_processing}.

\section{Experiments}\label{sec:experiments}

We apply the methods of Section~\ref{sec:methods} on five human-evaluation datasets, comparing against audio-only and model-intrinsic baselines on held-out test splits along the pipeline shown in Figure~\ref{fig:flowchart_models}.

\subsection{Datasets}
\label{subsec:exp_datasets}

We use five public human-evaluation benchmarks. MusicEval~\cite{liu2025musiceval} and SongEval~\cite{yao2025songeval} release per-clip $1$--$5$ expert ratings; MusicPref~\cite{huang2025musicpref}, AIME~\cite{grotschla2025aime}, and Music Arena~\cite{kim2025musicarena} release pairwise win/loss/tie outcomes (Music Arena adds a both-bad option). We keep only the musicality dimension---Musical Impression, Musicality, Music Quality, Musicality, and Music Arena's overall-preference vote, respectively---and drop orthogonal axes (text--audio alignment, fidelity, vocal naturalness, song-structure clarity). Splits are train/val/test with fixed seeds, stratified by generator system. Together the five benchmarks span ${\approx}241$\,h of rater-aligned audio: MusicEval ($2{,}748$ clips, ${\approx}17$\,h), SongEval ($2{,}399$ clips, ${\approx}140$\,h), AIME ($1{,}300$ clips, ${\approx}3.6$\,h), MusicPref ($5{,}040$ clips, ${\approx}20$\,h), and Music Arena ($6{,}080$ clips, ${\approx}61$\,h).

\subsection{Data Processing}
\label{subsec:exp_data_processing}

\textbf{Per-clip targets.}
MusicEval and SongEval provide continuous ratings directly: we average the musicality axis across annotators ($\sim\!5$ experts for MusicEval, $4$ for SongEval) to obtain $y_i\in[1,5]$. For the three pairwise benchmarks we apply the Elo scheme of Section~\ref{subsec:objective}~\cite{elo1978rating}. AIME exposes $12$ comparisons per clip, enough to fit per-clip Elo directly. MusicPref and Music Arena expose only one battle per clip, so we first fit Elo at the system level (each generator plays many matches) and assign each clip its system rating plus a single-match adjustment $K\,(S-E)$, avoiding collapse into three win/tie/loss bins.  Here $S$ is the observed score (win $1$, tie $0.5$, loss $0$; Music Arena's both-bad enters as $-0.5$ for both clips, ranking it below a loss), $E$ the rating-implied expectation, and $K$ the update step ($24$ at the system level, $16$ per clip). All Elo ratings are robust-affinely mapped to the shared
$1$--$5$ scale. This rescoring preserves ordinal content: per-clip/system Elo tracks empirical win rate at $r{\geq}0.86$ across all three benchmarks.

\textbf{Clip windowing.}
Feature extraction must operate on exactly the audio each rater heard. MusicEval ($\sim\!30$\,s) and MusicPref ($\sim\!10$\,s) use the native released clip. AIME uses the rater-aligned $10$-s window. SongEval rates full songs: we use the released audio under a $360$-s cap (trimming $0.28\%$ of the audio). For Music Arena we keep audio up to the shortest of reported listening time, released duration, and $180$\,s, dropping raters under $2$\,s and systems with fewer than $60$ clips ($6{,}080$ remain). Both corpora are then split into non-overlapping $30$-s chunks; we concatenate the per-chunk loss / entropy / SAE feature streams and uniformly mean-pool along time to $T{=}1500$ frames. The $30$-s chunk matches the $30$-s / $1500$-token context of MusicGen-Small~\cite{copet2023musicgen}; $T{=}1500$ is the fixed CNN input length (Section~\ref{subsec:exp_training}). Loss
and entropy curves use four parallel codebook channels plus a binary validity mask $m_t\in\{0,1\}$, giving $(5,T)$ per curve and $(9,T)$ when both are stacked.

\subsection{Model Training}
\label{subsec:exp_training}

\textbf{Architecture.}
Loss and entropy are from MusicGen-Small (teacher-forced); SAE codes are layer-$12$ activations via Singh et al.~\cite{singh2025steering} ($d_{\text{in}}{=}1024$, $d_{\text{sae}}{=}4096$).  A 1D CNN maps the stacked $(9,1500)$ loss/entropy curves (four codebook channels each plus mask) to $\mathbb{R}^{128}$; a parallel path compresses SAE codes $4096{\to}256$, applies two stride-$2$ Conv1d blocks, and pools to $\mathbb{R}^{128}$.  A two-layer MLP on the concatenation (Eq.~\eqref{eq:hybrid_concat}) yields $\hat y\in[1,5]$.  The full \textsc{Loss+Entropy+SAE} hybrid has $\approx\!1.24$\,M parameters ($\sim\!84\%$ in the SAE compression layer); single- and two-signal variants reuse the same encoders.

\textbf{Configurations.}
For every benchmark we train the seven feature combinations in Table~\ref{tab:main_results}---the three single-signal variants, the three pairwise hybrids, and the full \textsc{Loss+Entropy+SAE} hybrid---under identical splits, preprocessing and seeds, so cross-row comparisons isolate the effect of the input signals.  Training minimizes $\mathcal{L}_{\text{reg}}$ (Eq.~\eqref{eq:lreg}) with Adam (lr~$10^{-4}$, batch~$8$--$64$) and early-stops on the best-validation checkpoint.

\begin{figure}[t]
  \centering
  \includegraphics[width=\linewidth]{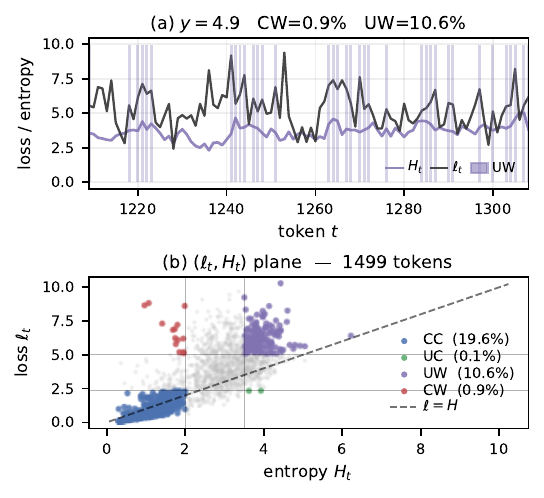}
  \caption{Example clip ($y{=}4.9$).  \textbf{(a)} 100-token window
    with the densest co-occurrence of CW and UW tokens.
    \textbf{(b)} Full-length $(\ell_t, H_t)$ plane for the same
    clip.}
  \label{fig:bidding}
\end{figure}
\subsection{Baselines}
\label{subsec:exp_baselines}

We compare the hybrid neural network of Section~\ref{subsec:architectures} and its single- and two-signal ablations against two reference groups. The first is \emph{audio-only aesthetics}: every test clip is scored by Meta Audiobox Aesthetics on its four axes---Content Enjoyment (CE), Content Usefulness (CU), Production Complexity (PC), Production Quality (PQ)---with a per-axis affine map $\hat z_i = \mathrm{clip}(a z_i + b,\,1,\,5)$ fit on validation only. The second, \emph{mean loss}, regresses the human rating on the clip's mean token loss.

\subsection{Results}
\label{subsec:exp_results}

We report Pearson $r$ (linear agreement on the $1$--$5$ scale) and Spearman $\rho$ (rank agreement) between predicted and human scores. Trained models use each benchmark's held-out test split; Audiobox is zero-shot on the full annotated set; and the \emph{All benchmarks} column merges the five held-out splits into one evaluation set. Within each column and model size, we \textbf{bold} the best score and \underline{underline} those statistically tied with it (paired clip-level bootstrap, $p{\geq}0.05$~\cite{demsar2006}).

Every intrinsic model beats the Audiobox aesthetics axes and the mean-loss baseline on the same clips, with \textsc{Single-SAE} the strongest single source except on MusicPref, where loss leads. Loss and entropy are individually predictive but add little beyond SAE: the curves' temporal structure, not a scalar summary, carries the signal---without ever processing the audio.

Combining everything is not always best. Grouping models by significance in each column (paired clip-level bootstrap~\cite{demsar2006}; bold/underline in Table~\ref{tab:main_results}), every benchmark has a top group set significantly apart from the rest, yet that group usually holds several tied models: its single best beats the runner-up in only $4$ of $24$ columns---three on the small model, one on the large.  The group's makeup shifts by benchmark---SAE-based models lead on most, while on MusicPref loss and entropy lead instead---and \textsc{Single-SAE} reaches the top group on most benchmarks, more consistently at large scale.  SAE carries most of the signal, and no combination dominates.

\section{Analysis}\label{sec:analysis}

The predictive relation above is promising; here we examine more empirically how loss, entropy, and the SAE latents relate to human ratings (MusicEval clean test split, $n{=}270$ clips). We first relate per-token combinations of loss and entropy to the ratings (Section~\ref{subsec:calibration}).  We then relate the spectral band power of the two curves to the ratings (Section~\ref{subsec:temporal}). Finally, we relate the SAE latents to the ratings through their interpretable concepts (Section~\ref{subsec:sae}).

\subsection{Loss--Entropy Configurations}
\label{subsec:calibration}

\begin{table}[t]
  \centering
  \small
  \setlength{\tabcolsep}{4pt}
  \begin{tabular}{lcc}
    \hline
    per-clip descriptor & $r$ & $\rho$ \\
    \hline
    $\tilde\pi_{\mathrm{CC}}$  (confident-correct)     & $+0.03$                & $+0.02$ \\
    $\tilde\pi_{\mathrm{UC}}$  (uncertain-correct)     & $-0.12^{\dagger}$      & $-0.12^{\dagger}$ \\
    $\tilde\pi_{\mathrm{CW}}$  (overconfident-wrong)   & $-0.18^{\ddagger}$     & $-0.19^{\ddagger}$ \\
    $\tilde\pi_{\mathrm{UW}}$  (uncertain-wrong)       & $+0.29^{\maltese}$ & $+0.32^{\maltese}$ \\
    $\tilde\pi_{\mathrm{UW}}-\tilde\pi_{\mathrm{CW}}$ & $+0.25^{\maltese}$ & $+0.27^{\maltese}$ \\
    $\overline{\ell-H}$  (excess surprise)             & $+0.30^{\maltese}$ & $+0.35^{\maltese}$ \\
    \hline
  \end{tabular}
  \caption{Per-clip loss--entropy configurations vs.\ human ratings
    (Pearson $r$, Spearman $\rho$).
    $^{\dagger}\,p{<}0.05$;
    $^{\ddagger}\,p{<}0.01$;
    $^{\maltese}\,p{<}0.001$; unmarked: $p{\geq}0.05$.}
  \label{tab:calibration_regimes}
\end{table}

Following Huron's probabilistic account of musical anticipation~\cite{huron2006sweet}, we read each generation step as an expectation and its outcome: $H_t$ measures how concentrated the prediction is, and $\ell_t$ measures how correct it turns out to be; for a categorical distribution these are the usual entropy / cross-entropy pair~\cite{cover2006elements}. Splitting each clip at its own $\ell_t$ and $H_t$ quartiles ($Q_{25}/Q_{75}$) labels each token with one of four configurations---confident-correct (CC: low $\ell$, low $H$), uncertain-correct (UC: low $\ell$, high $H$), overconfident-wrong (CW: high $\ell$, low $H$), and uncertain-wrong (UW: high $\ell$, high $H$); tokens inside the central interquartile band are unlabeled.  For each clip, $\tilde\pi_q$ is the proportion of its labeled tokens that fall in configuration $q$. Figure~\ref{fig:bidding} illustrates the four configurations on a highly rated clean example.

Across the MusicEval clean test split ($n{=}270$), three findings emerge (Table~\ref{tab:calibration_regimes}). \textbf{Overconfident-Wrong} ($\tilde\pi_{\mathrm{CW}}$) depresses rating ($r{=}{-}0.18$, $p{<}0.01$): the model's distribution was narrow, yet the actual token was unlikely under it---a local signature of audible distortion. \textbf{Uncertain-Wrong} ($\tilde\pi_{\mathrm{UW}}$) lifts rating ($r{=}{+}0.29$, $p{<}0.001$): even when the predictive distribution is already flat, the loss remains high, so these are moments that the model could not predict but a listener accepts. \textbf{Uncertain-Correct} ($\tilde\pi_{\mathrm{UC}}$) reverses sign ($r{=}{-}0.12$, $p{<}0.05$): the model is uncertain yet still correct; such moments are aimless and generic, and clips containing more of them are rated slightly lower. \textbf{Confident-Correct} ($\tilde\pi_{\mathrm{CC}}$) shows no correlation ($r{=}{+}0.03$, n.s.): what predicts the rating is not how often the model is right, but how it is wrong. A blind Gemini-2.5-Flash audit corroborates: $6/10$ CW windows are audibly disruptive vs.\ $3/10$ UW. The contrast $\tilde\pi_{\mathrm{UW}}{-}\tilde\pi_{\mathrm{CW}}$ collapses both effects into a single scalar ($r{=}{+}0.25$, $p{<}0.001$), and the simplest clip-level summary, $\overline{\ell\!-\!H}^{(i)}=\tfrac{1}{T_i}\sum_t(\ell_t-H_t)$ \cite{cover2006elements}, reaches $r{=}{+}0.30$ ($p{<}0.001$): higher-rated clips are systematically more surprising than the model's own uncertainty predicts.

\begin{figure}[t]
  \centering
  \includegraphics[width=\linewidth]{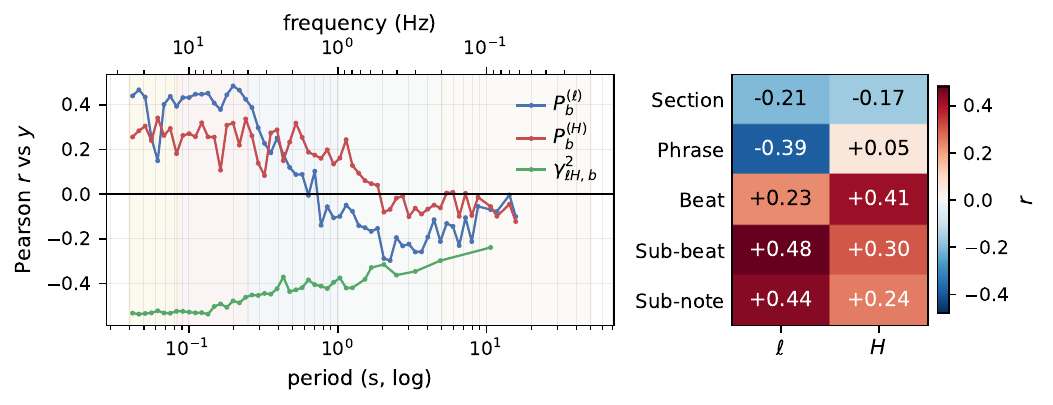}
  \caption{Spectral correlates of rating: Pearson $r$ between per-band
    power of $\ell_t,H_t$ and $y$ across log-spaced rate bands.
    \textbf{Left:} $64$ fine bands.  \textbf{Right:} five-band roll-up.}
  \label{fig:period_corr}
\end{figure}

\subsection{Loss and Entropy Signals in the Spectral Domain}
\label{subsec:temporal}

Section~\ref{subsec:calibration} summarized each clip by the distribution of $(\ell_t,H_t)$ values; here we analyze the curves' temporal dynamics.  We apply an FFT to each curve $s_t\in\{\ell_t,H_t\}$, writing $F_s(k)$ for the Fourier magnitude at frequency bin $k$, and define the relative band power $P_b^{(s)} = \sum_{k\in b}|F_s(k)|^2 \big/ \sum_k |F_s(k)|^2$ and $\log(1{+}P_b^{(s)})$ is the per-clip regressor against rating $y$. Bins are log-spaced between $0.05$ and $25$~Hz, rolled up into five named bands (\emph{Section} $\sim$5--50\,s, \emph{Phrase} $\sim$1--5\,s, \emph{Beat} $\sim$0.25--1\,s, \emph{Sub-beat} $\sim$80--250\,ms, \emph{Sub-note} $\sim$40--80\,ms).  In addition we estimate the magnitude-squared coherence $\gamma^2_{\ell H}(f)\in[0,1]$ between the two curves using Welch's averaged-periodogram method~\cite{welch1967spectra}. Fig.~\ref{fig:period_corr} reports the per-band Pearson $r$ against $y$ (left: $64$ fine bands; right: the five named bands).

Three musical patterns emerge. (i)~\textbf{Rapid phrase-rate change aligns with lower ratings} ($r{\approx}{-}0.39$ in the \emph{Phrase} band; Fig.~\ref{fig:period_corr} right): a surprise track that keeps re-starting every few seconds reads as formally unstable. (ii)~\textbf{Beat-rate variation aligns with higher ratings} ($r{\approx}{+}0.41$ for $H$ at the beat, $r{\approx}{+}0.48$ for $\ell$ at the sub-beat; Fig.~\ref{fig:period_corr} right): a clear rhythmic pulse matches perceived groove. (iii)~\textbf{Simultaneous changes in loss and entropy align with lower ratings} ($\gamma^2$: $r{\approx}{-}0.54$ across $40$--$190$~ms; Fig.~\ref{fig:period_corr} left): at the millisecond grain this is the signature of glitchy audio, where the model is jointly confused and wrong on every micro-event.

\subsection{SAE Latents as Interpretable Concepts}
\label{subsec:sae}

To verify that the SAE predictions rest on audible musical traits rather than dataset artifacts, we compute for each SAE latent $k$ a per-clip gradient attribution $A_k^{(i)}=\sum_t (\partial \hat{y}_i/\partial z_t^{(i,k)})\,z_t^{(i,k)}$ from the trained Single-SAE model and rank latents by $\rho_k=\mathrm{Pearson}(A_k^{(\cdot)},\,y)$ ($365/4096$ latents admit a finite $\rho_k$). The top-activating $2$-s windows of three groups---good-aligned ($10$ largest $+\rho_k$), bad-aligned ($10$ largest $-\rho_k$), and near-null ($|\rho_k|\!<\!0.01$, $5$ per sign)---are captioned by Gemini-$2.5$-Flash~\cite{google2025gemini25} and, without seeing these captions or $\rho_k$, by a professional music producer (Table~\ref{tab:sae_concept_labels}).\footnote{The annotating producer has worked professionally for six years across electronic, pop, and orchestral production.} On the good side, the producer and the VLM agree on conventional musical content; on the bad side the VLM still issues clean genre labels while the producer reaches for distortion terms (\emph{noise}, \emph{harsh hi-freq}, \emph{EOS/delay tokens}).  The producer's judgment matches the $\rho_k$ ranking, so gradient attribution recovers the same quality axis a trained listener would.

\begin{table}[t]
  \centering
  \scriptsize
  \setlength{\tabcolsep}{3pt}
  \begin{tabular}{@{}lcp{1.9cm}@{\hspace{0.6mm}}p{2.8cm}@{}}
    \hline
    rank & $\rho_k$ & VLM caption & producer concept \\
    \hline
    \multicolumn{4}{l}{\emph{good-aligned} ($\rho_k>0$)} \\
    g-1   & \rhobar{+0.54}   & classical, string         & string classical \\
    g-2   & \rhobar{+0.49}   & folk, upbeat              & folk guitar texture \\
    g-5   & \rhobar{+0.45}   & folk, calm                & choir + plucked folk \\
    g-7   & \rhobar{+0.44}   & folk, melancholic         & flute texture \\
    g-9   & \rhobar{+0.43}   & ambient, calm             & ambient pads + vocal tex.\\
    g-10  & \rhobar{+0.43}   & pop, upbeat               & clear pop with vocal+beat \\
    \hline
    \multicolumn{4}{l}{\emph{near-null} ($|\rho_k|<0.01$)} \\
    m$_+$3 & \rhobar{+0.005} & rock, energetic           & loud rock, single long note \\
    m$_+$1 & \rhobar{+0.001} & pop, energetic            & clear melody+accomp. \\
    m$_-$1 & \rhobar{-0.001} & classical, melanch.       & harsh sound, bad groove \\
    m$_-$3 & \rhobar{-0.002} & hip-hop, energetic        & bad groove, noise \\
    \hline
    \multicolumn{4}{l}{\emph{bad-aligned} ($\rho_k<0$)} \\
    b-9   & \rhobar{-0.43}   & classical, playful        & dissonant wide-range \\
    b-8   & \rhobar{-0.45}   & --- (silence)             & EOS / delay tokens \\
    b-7   & \rhobar{-0.46}   & classical, melanch.       & dissonant violin texture \\
    b-5   & \rhobar{-0.47}   & hip-hop, tense            & random perc., harsh hi-freq \\
    b-4   & \rhobar{-0.48}   & pop, upbeat               & bad vocal+noisy perc.\\
    b-3   & \rhobar{-0.48}   & --- (silence)             & EOS / delay tokens \\
    b-2   & \rhobar{-0.60}   & rock, energetic           & big noise, random trumpet \\
    b-1   & \rhobar{-0.61}   & classical, calm           & random hi-freq noise \\
    \hline
  \end{tabular}
  \caption{Representative SAE latents ranked by rating-aligned
    attribution $\rho_k$, each annotated by Gemini-$2.5$-Flash and by
    a professional music producer. Thin red/blue bars in the $\rho_k$
    column visualize $|\rho_k|$.}
  \label{tab:sae_concept_labels}
\end{table}

\section{Conclusion}\label{sec:conclusion}

A music generative model's own intrinsic signals already track which of its outputs listeners rate highly. Loss, entropy, and SAE latents from a single frozen MusicGen align with human ratings on five public benchmarks and outperform audio-only aesthetics baselines: loss--entropy mismatch flags the overconfident-wrong moments that disrupt perceived quality, and SAE latents organize clips along a quality axis without ever seeing a rating. Intrinsic signals are thus a viable, underused axis for automatic music evaluation.

\section{Acknowledgments}

We thank Professor Alexander Lerch for his guidance and his review
of this paper.

\bibliography{ISMIRtemplate}

@ARTICLE{lerch2025survey,
        AUTHOR = {Alexander Lerch and Claire Arthur and Nick Bryan-Kinns and Corey Ford and Qianyi Sun and Ashvala Vinay},
        TITLE = {Survey on the Evaluation of Generative Models in Music},
        JOURNAL = {ACM Computing Surveys},
        VOLUME = {58},
        NUMBER = {4},
        YEAR = {2025}
}

@ARTICLE{tjandra2025audiobox,
        AUTHOR = {Andros Tjandra and Yi-Chiao Wu and Baishan Guo and John Hoffman and Brian Ellis and Apoorv Vyas and Bowen Shi and Sanyuan Chen and Matt Le and Nick Zacharov and Carleigh Wood and Ann Lee and Wei-Ning Hsu},
        TITLE = {Meta Audiobox Aesthetics: Unified Automatic Quality Assessment for Speech, Music, and Sound},
        JOURNAL = {arXiv preprint arXiv:2502.05139},
        YEAR = {2025}
}

@INPROCEEDINGS{ma2025cmibench,
        AUTHOR = {Yinghao Ma and Siyou Li and Juntao Yu and Emmanouil Benetos and Akira Maezawa},
        TITLE = {{CMI-Bench}: A Comprehensive Benchmark for Evaluating Music Instruction Following},
        BOOKTITLE = {Proc. of the 26th Int. Society for Music Information Retrieval Conf. (ISMIR)},
        PAGES = {416--425},
        YEAR = {2025}
}

@ARTICLE{ma2026cmireward,
        AUTHOR = {Yinghao Ma and Haiwen Xia and Hewei Gao and Weixiong Chen and Yuxin Ye and Yuchen Yang and Sungkyun Chang and Mingshuo Ding and Yizhi Li and Ruibin Yuan and Simon Dixon and Emmanouil Benetos},
        TITLE = {{CMI-RewardBench}: Evaluating Music Reward Models with Compositional Multimodal Instruction},
        JOURNAL = {arXiv preprint arXiv:2603.00610},
        YEAR = {2026}
}

@INPROCEEDINGS{liu2025musiceval,
        AUTHOR = {Cheng Liu and Hui Wang and Jinghua Zhao and Shiwan Zhao and Hui Bu and Xin Xu and Jiaming Zhou and Haoqin Sun and Yong Qin},
        TITLE = {{MusicEval}: A Generative Music Dataset with Expert Ratings for Automatic Text-to-Music Evaluation},
        BOOKTITLE = {Proc. of the IEEE Int. Conf. on Acoustics, Speech and Signal Processing (ICASSP)},
        PAGES = {1--5},
        YEAR = {2025}
}

@ARTICLE{yao2025songeval,
        AUTHOR = {Jixun Yao and Guobin Ma and Huixin Xue and Huakang Chen and Chunbo Hao and Yuepeng Jiang and Haohe Liu and Ruibin Yuan and Jin Xu and Wei Xue and Hao Liu and Lei Xie},
        TITLE = {{SongEval}: A Benchmark Dataset for Song Aesthetics Evaluation},
        JOURNAL = {arXiv preprint arXiv:2505.10793},
        YEAR = {2025}
}

@INPROCEEDINGS{huang2025musicpref,
        AUTHOR = {Yichen Huang and Zachary Novack and Koichi Saito and Jiatong Shi and Shinji Watanabe and Yuki Mitsufuji and John Thickstun and Chris Donahue},
        TITLE = {Aligning Text-to-Music Evaluation with Human Preferences},
        BOOKTITLE = {Proc. of the 26th Int. Society for Music Information Retrieval Conf. (ISMIR)},
        YEAR = {2025}
}

@INPROCEEDINGS{grotschla2025aime,
        AUTHOR = {Florian Gr{\"o}tschla and Ahmet Solak and Luca A. Lanzend{\"o}rfer and Roger Wattenhofer},
        TITLE = {Benchmarking Music Generation Models and Metrics via Human Preference Studies},
        BOOKTITLE = {Proc. of the IEEE Int. Conf. on Acoustics, Speech, and Signal Processing (ICASSP)},
        YEAR = {2025}
}

@INPROCEEDINGS{kim2025musicarena,
        AUTHOR = {Yonghyun Kim and Wayne Chi and Anastasios N. Angelopoulos and Wei-Lin Chiang and Koichi Saito and Shinji Watanabe and Yuki Mitsufuji and Chris Donahue},
        TITLE = {Music Arena: Live Evaluation for Text-to-Music},
        BOOKTITLE = {Advances in Neural Information Processing Systems 38 (NeurIPS), Creative AI Track},
        YEAR = {2025}
}

@INPROCEEDINGS{chiang2024chatbotarena,
        AUTHOR = {Wei-Lin Chiang and Lianmin Zheng and Ying Sheng and Anastasios Nikolas Angelopoulos and Tianle Li and Dacheng Li and Banghua Zhu and Hao Zhang and Michael I. Jordan and Joseph E. Gonzalez and Ion Stoica},
        TITLE = {Chatbot Arena: An Open Platform for Evaluating {LLM}s by Human Preference},
        BOOKTITLE = {Proc. of the 41st Int. Conf. on Machine Learning (ICML)},
        PAGES = {8359--8388},
        YEAR = {2024}
}

@ARTICLE{li2026noiseloss,
        AUTHOR = {Xiaosha Li and Chun Liu and Ziyu Wang},
        TITLE = {When Noise Lowers the Loss: Rethinking Likelihood-Based Evaluation in Music Large Language Models},
        JOURNAL = {arXiv preprint arXiv:2602.02738},
        YEAR = {2026}
}

@ARTICLE{velickovic2026perplexity,
        AUTHOR = {Petar Veli{\v{c}}kovi{\'c} and Federico Barbero and Christos Perivolaropoulos and Simon Osindero and Razvan Pascanu},
        TITLE = {Perplexity Cannot Always Tell Right from Wrong},
        JOURNAL = {arXiv preprint arXiv:2601.22950},
        YEAR = {2026}
}

@INPROCEEDINGS{cunningham2023sparse,
        AUTHOR = {Hoagy Cunningham and Aidan Ewart and Logan Riggs and Robert Huben and Lee Sharkey},
        TITLE = {Sparse Autoencoders Find Highly Interpretable Features in Language Models},
        BOOKTITLE = {Int. Conf. on Learning Representations (ICLR)},
        YEAR = {2024}
}

@MISC{bricken2023monosemanticity,
        AUTHOR = {Trenton Bricken and Adly Templeton and Joshua Batson and Brian Chen and Adam Jermyn and Tom Conerly and Nick Turner and Cem Anil and Carson Denison and Amanda Askell and Robert Lasenby and Yifan Wu and Shauna Kravec and Nicholas Schiefer and Tim Maxwell and Nicholas Joseph and Zac Hatfield-Dodds and Alex Tamkin and Karina Nguyen and Brayden McLean and Josiah E Burke and Tristan Hume and Shan Carter and Tom Henighan and Christopher Olah},
        TITLE = {Towards Monosemanticity: Decomposing Language Models with Dictionary Learning},
        HOWPUBLISHED = {Transformer Circuits Thread},
        YEAR = {2023},
        URL = {https://transformer-circuits.pub/2023/monosemantic-features}
}

@MISC{templeton2024scaling,
        AUTHOR = {Adly Templeton and Tom Conerly and Jonathan Marcus and Jack Lindsey and Trenton Bricken and Brian Chen and Adam Pearce and Craig Citro and Emmanuel Ameisen and Andy Jones and Hoagy Cunningham and Nicholas L Turner and Callum McDougall and Monte MacDiarmid and C. Daniel Freeman and Theodore R. Sumers and Edward Rees and Joshua Batson and Adam Jermyn and Shan Carter and Chris Olah and Tom Henighan},
        TITLE = {Scaling Monosemanticity: Extracting Interpretable Features from {Claude 3 Sonnet}},
        HOWPUBLISHED = {Transformer Circuits Thread},
        YEAR = {2024},
        URL = {https://transformer-circuits.pub/2024/scaling-monosemanticity}
}

@INPROCEEDINGS{singh2025steering,
        AUTHOR = {Nikhil Singh and Manuel Cherep and Pattie Maes},
        TITLE = {Discovering and Steering Interpretable Concepts in Large Generative Music Models},
        BOOKTITLE = {Int. Conf. on Learning Representations (ICLR)},
        YEAR = {2026}
}

@BOOK{elo1978rating,
        AUTHOR = {Arpad E. Elo},
        TITLE = {The Rating of Chessplayers, Past and Present},
        PUBLISHER = {Arco Publishing},
        ADDRESS = {New York},
        YEAR = {1978}
}

@ARTICLE{bradley1952rank,
        AUTHOR = {Ralph Allan Bradley and Milton E. Terry},
        TITLE = {Rank Analysis of Incomplete Block Designs: {I}. The Method of Paired Comparisons},
        JOURNAL = {Biometrika},
        VOLUME = {39},
        NUMBER = {3/4},
        PAGES = {324--345},
        YEAR = {1952}
}

@INPROCEEDINGS{burges2005learning,
        AUTHOR = {Christopher Burges and Tal Shaked and Erin Renshaw and Ari Lazier and Matt Deeds and Nicole Hamilton and Greg Hullender},
        TITLE = {Learning to Rank Using Gradient Descent},
        BOOKTITLE = {Proceedings of the 22nd International Conference on Machine Learning (ICML)},
        PAGES = {89--96},
        YEAR = {2005}
}

@INPROCEEDINGS{herbrich1999support,
        AUTHOR = {Ralf Herbrich and Thore Graepel and Klaus Obermayer},
        TITLE = {Support Vector Learning for Ordinal Regression},
        BOOKTITLE = {Proceedings of the 9th International Conference on Artificial Neural Networks (ICANN)},
        PAGES = {97--102},
        YEAR = {1999}
}

@MISC{google2025gemini25,
        AUTHOR = {{Gemini Team, Google}},
        TITLE = {{Gemini 2.5}: Pushing the Frontier with Advanced Reasoning, Multimodality, Long Context, and Next Generation Agentic Capabilities},
        HOWPUBLISHED = {arXiv preprint arXiv:2507.06261},
        YEAR = {2025}
}

@BOOK{huron2006sweet,
        AUTHOR = {David Huron},
        TITLE = {Sweet Anticipation: Music and the Psychology of Expectation},
        PUBLISHER = {MIT Press},
        ADDRESS = {Cambridge, MA},
        YEAR = {2006},
        ISBN = {9780262083454}
}

@BOOK{cover2006elements,
        AUTHOR = {Thomas M. Cover and Joy A. Thomas},
        TITLE = {Elements of Information Theory},
        EDITION = {2nd},
        PUBLISHER = {Wiley},
        YEAR = {2006}
}

@INPROCEEDINGS{guo2017calibration,
        AUTHOR = {Chuan Guo and Geoff Pleiss and Yu Sun and Kilian Q. Weinberger},
        TITLE = {On Calibration of Modern Neural Networks},
        BOOKTITLE = {Proc. of the 34th Int. Conf. on Machine Learning (ICML)},
        PAGES = {1321--1330},
        YEAR = {2017}
}

@INPROCEEDINGS{pereyra2017confidence,
        AUTHOR = {Gabriel Pereyra and George Tucker and Jan Chorowski and {\L}ukasz Kaiser and Geoffrey E. Hinton},
        TITLE = {Regularizing Neural Networks by Penalizing Confident Output Distributions},
        BOOKTITLE = {Int. Conf. on Learning Representations (ICLR), Workshop Track},
        YEAR = {2017}
}

@ARTICLE{qiu2024entropy,
        AUTHOR = {Zexuan Qiu and Zijing Ou and Bin Wu and Jingjing Li and Aiwei Liu and Irwin King},
        TITLE = {Entropy-Based Decoding for Retrieval-Augmented Large Language Models},
        JOURNAL = {arXiv preprint arXiv:2406.17519},
        YEAR = {2024}
}

@INPROCEEDINGS{malinin2021uncertainty,
        AUTHOR = {Andrey Malinin and Mark Gales},
        TITLE = {Uncertainty Estimation in Autoregressive Structured Prediction},
        BOOKTITLE = {Int. Conf. on Learning Representations (ICLR)},
        YEAR = {2021}
}

@ARTICLE{fomicheva2020unsupervised,
        AUTHOR = {Marina Fomicheva and Shuo Sun and Lisa Yankovskaya and Fr{\'e}d{\'e}ric Blain and Ale{\v{s}} Tamchyna and Mark Fishel and Nikola Popovi{\'c} and M{\'o}nica L{\'o}pez and Vishrav Chaudhary and Angela Fan and Yvette Graham and Barry Haddow and Matthias Huck and Antonio N. Valentino and Marco Turchi and Lucia Specia},
        TITLE = {Unsupervised Quality Estimation for Neural Machine Translation},
        JOURNAL = {Transactions of the Association for Computational Linguistics},
        VOLUME = {8},
        PAGES = {639--657},
        YEAR = {2020}
}

@BOOK{meyer1956emotion,
        AUTHOR = {Leonard B. Meyer},
        TITLE = {Emotion and Meaning in Music},
        PUBLISHER = {University of Chicago Press},
        ADDRESS = {Chicago, IL},
        YEAR = {1956}
}

@ARTICLE{zajonc1968mereexposure,
        AUTHOR = {Robert B. Zajonc},
        TITLE = {Attitudinal Effects of Mere Exposure},
        JOURNAL = {Journal of Personality and Social Psychology},
        VOLUME = {9},
        NUMBER = {2, Pt.2},
        PAGES = {1--27},
        YEAR = {1968}
}

@ARTICLE{schellenberg2008exposure,
        AUTHOR = {E. Glenn Schellenberg and Isabelle Peretz and Sandra Vieillard},
        TITLE = {Liking for Happy- and Sad-Sounding Music: Effects of Exposure},
        JOURNAL = {Cognition and Emotion},
        VOLUME = {22},
        NUMBER = {2},
        PAGES = {218--237},
        YEAR = {2008}
}

@ARTICLE{welch1967spectra,
        AUTHOR = {P. Welch},
        TITLE = {The Use of Fast {Fourier} Transform for the Estimation of Power Spectra: {A} Method Based on Time Averaging Over Short, Modified Periodograms},
        JOURNAL = {IEEE Transactions on Audio and Electroacoustics},
        VOLUME = {15},
        NUMBER = {2},
        PAGES = {70--73},
        YEAR = {1967}
}

@INPROCEEDINGS{copet2023musicgen,
        AUTHOR = {Jade Copet and Felix Kreuk and Itai Gat and Tal Remez and David Kant and Gabriel Synnaeve and Yossi Adi and Alexandre D{\'e}fossez},
        TITLE = {Simple and Controllable Music Generation},
        BOOKTITLE = {Advances in Neural Information Processing Systems 36 (NeurIPS)},
        YEAR = {2023}
}

@INPROCEEDINGS{muellereberstein2023subspace,
        AUTHOR = {Max M{\"u}ller-Eberstein and Rob van der Goot and Barbara Plank and Ivan Titov},
        TITLE = {Subspace Chronicles: How Linguistic Information Emerges, Shifts and Interacts during Language Model Training},
        BOOKTITLE = {Findings of the Association for Computational Linguistics: EMNLP 2023},
        YEAR = {2023}
}

@ARTICLE{demsar2006,
        AUTHOR = {Janez Dem{\v{s}}ar},
        TITLE = {Statistical Comparisons of Classifiers over Multiple Data Sets},
        JOURNAL = {Journal of Machine Learning Research},
        VOLUME = {7},
        PAGES = {1--30},
        YEAR = {2006}
}

@INCOLLECTION{justus2002music,
        AUTHOR = {Timothy C. Justus and Jamshed J. Bharucha},
        TITLE = {Music Perception and Cognition},
        BOOKTITLE = {Stevens' Handbook of Experimental Psychology, Vol.~1: Sensation and Perception},
        PUBLISHER = {John Wiley \& Sons},
        PAGES = {453--492},
        YEAR = {2002}
}

@ARTICLE{koelsch2011toward,
        AUTHOR = {Stefan Koelsch},
        TITLE = {Toward a Neural Basis of Music Perception -- A Review and Updated Model},
        JOURNAL = {Frontiers in Psychology},
        VOLUME = {2},
        PAGES = {110},
        YEAR = {2011}
}

\end{document}